\documentclass[fleqn,10pt]{wlscirep}
\usepackage[utf8]{inputenc}
\usepackage[T1]{fontenc}
\usepackage{mathtools}
\usepackage{graphicx}
\usepackage{tikz}
\usepackage{svg}
\usepackage{wrapfig}
\usepackage{wrapfig}
\usepackage[demo, export]{adjustbox}
\usepackage{float, tabularx, booktabs}
\newcolumntype{Y}{>{\centering\arraybackslash}X}

\usepackage{subcaption}
\usepackage[utf8]{inputenc}
\usepackage{pgfplots}
\DeclareUnicodeCharacter{2212}{−}
\usepgfplotslibrary{groupplots,dateplot}
\usetikzlibrary{patterns,shapes.arrows}
\pgfplotsset{compat=newest}

\usepackage{xcolor}

\usepackage[normalem]{ulem}

\title{How Media Competition Fuels the Spread of Misinformation}

\author[1,*]{Arash Amini}
\author[2,*]{Yigit Ege Bayiz}
\author[3]{Eun-Ju Lee}
\author[4]{Zeynep Somer-Topcu}
\author[2]{Radu Marculescu}
\author[1]{Ufuk Topcu}

\affil[1]{The University of Texas at Austin, Oden Institute for Computational Engineering and Sciences}
\affil[2]{The University of Texas at Austin, Department of Electrical and Computer Engineering}
\affil[3]{Seoul National University, Department of Communication \& Interdisciplinary Program in Artificial Intelligence}
\affil[4]{The University of Texas at Austin, Department of Government}
\affil[*]{Equal Contribution}

\begin{abstract}

Competition among news sources may encourage some sources to share fake news and misinformation to influence the public.
While sharing misinformation may lead to a short-term gain in audience engagement, it may damage the reputation of these sources, resulting in a loss of audience.
To understand the rationale behind sharing misinformation, we model the competition as a zero-sum sequential game, where each news source influences individuals based on its credibility---how trustworthy the public perceives it---and the individual's opinion and susceptibility. 
In this game, news sources can decide whether to share factual information to enhance their credibility or disseminate misinformation for greater immediate attention at the cost of losing credibility. We employ the quantal response equilibrium concept, which accounts for the bounded rationality of human decision-making, allowing for imperfect or probabilistic choices. Our analysis shows that the resulting equilibria for this game reproduce the credibility-bias distribution observed in real-world news sources, with hyper-partisan sources more likely to spread misinformation than centrist ones. It further illustrates that disseminating misinformation can polarize the public. Notably, our model reveals that when one player increases misinformation dissemination, the other player is likely to follow, exacerbating the spread of misinformation. We conclude by discussing potential strategies to mitigate the spread of fake news and promote a more factual and reliable information landscape.

\end{abstract}
\begin{document}

\flushbottom
\maketitle

\thispagestyle{empty}

\section*{Introduction}

Controlling information offers a significant advantage in political and economic spheres. Consequently, it drives media corporations and political influencers to compete for dominance. Actors in this competition employ various strategies to gain a larger market share. Strategies include targeting specific groups or promoting inclusivity by presenting diverse viewpoints. However, some opt for deceptive tactics, such as disseminating misinformation, rumors, and sensational narratives, to generate interest and attract the attention of news audiences. \cite{vosoughi2018spread,graves2019information}. The employment of misinformation for political gain is historically well-documented. For instance, Octavian, the founder of the Roman Empire, employed propaganda to smear the reputations of his rivals by spreading fake stories  \cite{posetti2018short}. 

Contrary to the widespread belief that social media is the primary source of misinformation, research indicates that online fake news only reaches a limited audience. In contrast, the widespread exposure and internalization of misinformation among the general population underscore the significant role of public figures and traditional media channels in its dissemination \cite{tsfati2020causes}. Various factors contribute to this issue, including the inherent news value of fake news, which tends to attract attention  \cite{aimeur2023fake}.  In addition to the potentially devastating consequences of false information to society, from a purely strategic perspective, resorting to misinformation has a significant strategic drawback \cite{ognyanova2020misinformation}. Over time, as the public becomes increasingly wary of potential deceit, its response to information low credible sources share diminishes. Growing skepticism can lead many to perceive the news source as unreliable, resulting in a gradual loss of interest in its reporting.

In the past decade, researchers have increasingly focused on studying misinformation, particularly its production \cite{ecker2022psychological, tsfati2020causes,chen2013misinformation}, dissemination \cite{vosoughi2018spread, de2013anatomy, aimeur2023fake,bayiz2024susceptibility}, detection \cite{wu2019misinformation,kumar2014detecting,amini2024control,ram2024credirag,bayiz2024optimization}, and countering \cite{lewandowsky2021countering,roozenbeek2023countering,chan2017debunking,altay2023survey,bayiz2023prebunking}. However, the potential relationship between news source competition on public opinion and the {production} of misinformation remains poorly understood. In this paper, we adopt a distinct approach compared to existing literature. Instead of exploring a single instance of fake news and its spread through social networks \cite{vemprala2021debunking,roozenbeek2023countering}, we choose to study the long-term effects of frequent misinformation dissemination and its impact on the information landscape.

 The rise of social networks has transformed the information landscape\cite{van2016weapon}, challenging the monopoly of traditional media. Thus, incorporating social media into competition modeling between news sources is essential.  To model this competition, we employ \textit{opinion dynamics} modeling, which is widely accepted for studying information flow and the evolution of individual opinions over time under diverse interactions \cite{del2016spreading,acemoglu2010spread,acemoglu2011opinion}. To mathematically explain why news sources share misinformation, we introduce a novel misinformation model for media competition. News sources in our model share information with constant frequency. We assume that each source can decide whether to disseminate misinformation or factual information at each time instance. While we simplify the choices to binary, in reality, however, they exist on a spectrum---from completely misleading to factual.

\begin{figure}
    \centering
    \includegraphics[width=1\linewidth]{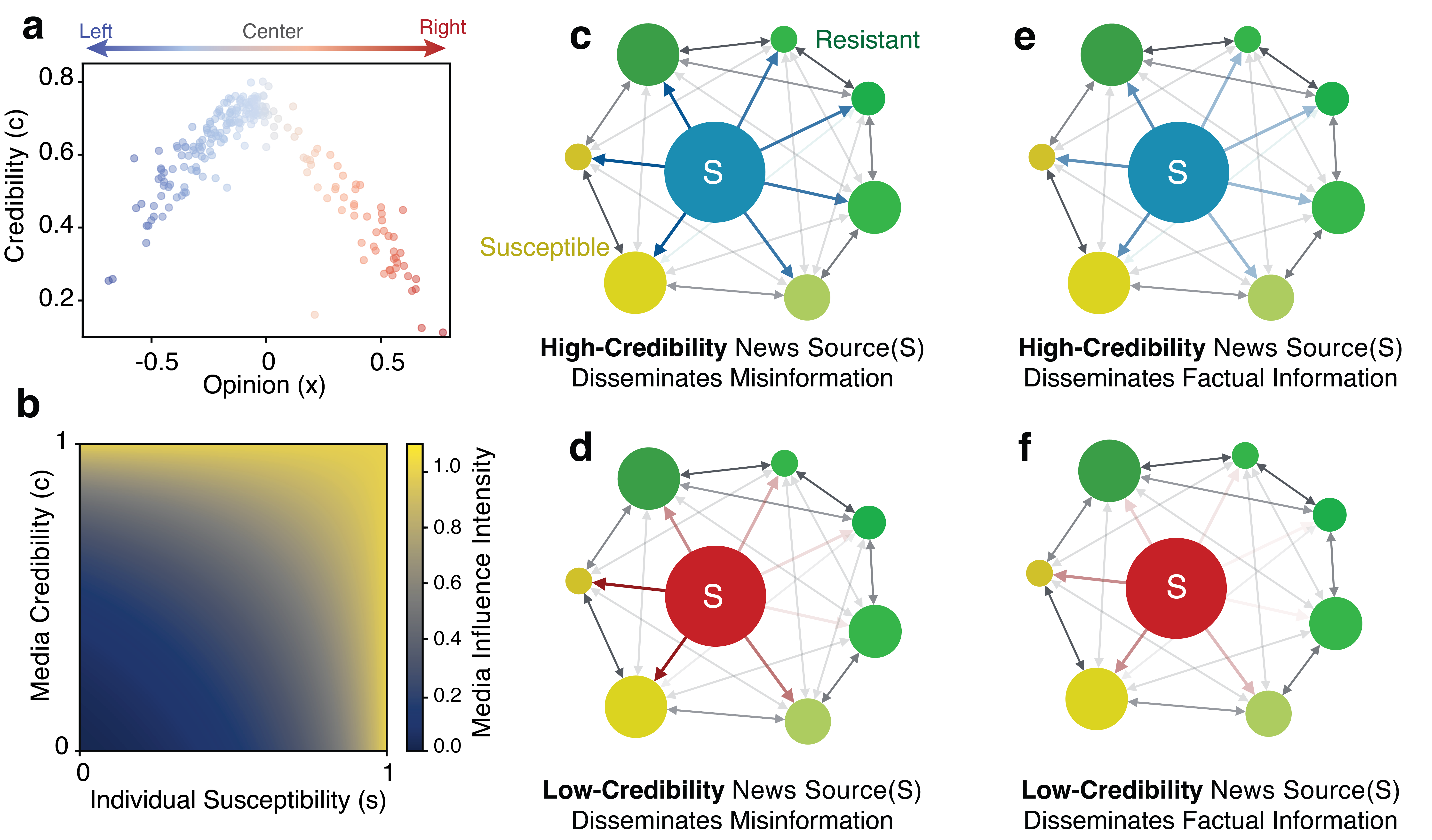}
    \caption{\textbf{a.} The credibility with respect to political bias for 223 news sources from Ad-Font websites. The hyper-partisan sources are less credible compared to the centrist ones. \textbf{b.} Normalized media influence intensity with respect to news source credibility and individual susceptibility. Credibility becomes irrelevant for susceptible individuals. \textbf{c-f.} Visualization of how credibility and disseminating misinformation affect individuals with different levels of susceptibility. Green users are resistant to misinformation, while yellow users are susceptible. The opacity of the arrows increases with increased influence. 
    \textbf{c,e.} Sources with high credibility (blue) significantly influence all users, even when disseminating misinformation. \textbf{d,f.} Sources with low credibility (red) have almost no influence over less susceptible individuals; however, disseminating misinformation substantially increases its influence over susceptible individuals.  }
    \label{fig:Ad-Font}
\end{figure}

We specifically hypothesize that fake news and misinformation possess an intrinsic news value, which attracts a broader audience\cite{tsfati2020causes, kurvers2021strategic}. We base the proposed model on the assumption that while the short-term attention gained from the dissemination of fake news can be significant, its prolonged circulation ultimately erodes community trust, leading the public to question the credibility of news sources\cite{lewandowsky2017beyond}.To model opinion dynamics in scenarios where news sources disseminate misinformation, we first formally define the credibility of a news source based on the frequency of fake news disseminated by a source. We then use the credibility metric and the types of information the source distributes to determine its influence on individuals.

We then assume that individuals are influnced by news based on their susceptibility, the chance of accepting information without regard for the source, and the objective credibility of the source.
We posit that people are influenced by news they receive according to their bias and susceptibility---The ability of individuals to critically assess both the origin and reliability of the news content with which they engage---as well as the source's objective credibility and bias. In other words, individuals are skeptical only of the source's credibility, not the content of the news being shared. As a result, even resistant (low-susceptible) individuals can be influenced by misinformation from credible sources. Although this assumption differs from the classical definition of susceptibility, it more accurately reflects the behavior of the general public in consuming news \cite{bhattacherjee2022effects,van2022misinformation}.

Finally, we investigate how competition between news sources fosters misinformation dissemination by modeling the competition as a zero-sum game. In this game, two players----each representing all ideologically similar sources---control information dissemination strategies for their respective sources. We investigate the quantal response equilibrium of the zero-sum game setup. This equilibrium is a generalization of the Nash equilibrium and incorporates a tunable additive entropy regularization term that represents the inherent randomness and bounded rationality in human decision-making. Our findings indicate that the equilibria highlight four critical phenomena. First, the equilibrium replicates the credibility-opinion distribution seen in real-world news sources (Figure \ref{fig:Ad-Font}), particularly encouraging hyperpartisan media to disseminate more misinformation than centrist ones. Second, it explains how dissemination of misinformation contributes to ideological opinion polarization and the formation of echo chambers. Third, it shows that when one player increases misinformation, the equilibrium forces the opposing player to do the same, amplifying the spread of misinformation. Finally, it demonstrates that while improving public susceptibility and penalizing the media for misinformation can reduce polarization, these measures may not necessarily lower overall community exposure to misinformation.

Our main contributions are threefold. 
\begin{itemize}
    \item First, we introduce a model that details the effects of news source credibility and misinformation dissemination on the evolution of public opinion, examining the long-term impact of misinformation exposure, and expanding the scope of the paper from singular instances of fake news spread to broader trends.
    \item Second, by examining the resulting equilibrium, we analyze how competitive dynamics among news sources encourage the frequent spread of misinformation. This analysis offers insights into how news sources determine the quality of information they distribute.
    \item Third, we evaluate potential interventions policy makers could implement to reduce ideological polarization and misinformation exposure in the public, using our findings on equilibrium.
\end{itemize}

For consistency, it is essential to establish a clear definition for the \textit{fake news}, \textit{misinformation}, and \textit{news source} since these terms have different meanings in various contexts. We define \textit{fake news} and \textit{misinformation} as any news or information that is partially or entirely incorrect, misleading, sensational and disseminated to manipulate the public. Our definition of a \textit{news source} includes persons (e.g. influencers on social media) and organizations (e.g., media corporations) that publish content frequently to a mass audience. This definition encompasses large media corporations and groups of individual influencers on social media who often share information with large audiences. Therefore, this paper employs \textit{news sources} rather than \textit{media} to encompass a broader range of news-sharing platforms.

The paper is structured to first introduce the model in which news sources can disseminate misinformation to gain additional attention at a cost to their credibility. We then examine the competition over public influence by examining the equilibrium. Finally, we conclude with a discussion on intervention strategies to foster a more balanced information environment and enhance public discourse.

\subsection*{Related Works}

\paragraph{Causes and Consequences of Misinformation}
The existing research on the causes of misinformation aims to identify the driving political\cite{pickard2016media}, social\cite{sindermann2020short}, and psychological\cite{pennycook2021psychology} reasons behind misinformation propagation. The oldest treatises discussing the political use of misinformation can be traced back to the renaissance. However, with widespread news sharing via social media and the abundance of online data, there has been a surge in recent research on the causes of misinformation. Misinformation propagates readily between users over online social networks \cite{aimeur2023fake, del2016spreading}. However, Tsfati et al. show that mainstream media remains also remains to be among the main propagators of misinformation \cite{tsfati2020causes}, necessitating the inclusion of mainstream media outlets' influence in misinformation models. Recent studies also highlight the role of political elites in community exposure to misinformation on online platforms \cite{lasser2022social,mosleh2022measuring}, and how it spreads \cite{vosoughi2018spread}. Understanding the mechanisms and motivations behind the spread of misinformation by different entities is still in its infancy \cite{tsfati2020causes}.  
To this end, we employ game theory in opinion dynamics models to better understand the stakeholders' decision-making and why they resort to sharing misinformation.

\textit{Reflexivity}, the ability of users to reflect on the source and accuracy of the news they consume, is one of the main factors in misinformation propagation that is more influential than partisanship \cite{pennycook2021psychology}. Our model incorporates reflexivity as an innate susceptibility parameter, which describes how much each user gets influenced by low credibility sources and misinformation. The innate susceptibility of users is the main driving factor in misinformation propagation in non-political topics such as COVID-19 rumors \cite{roozenbeek2020susceptibility,kouzy2020coronavirus}. Nevertheless, political leaning is still a factor in the propagation of politics-related misinformation, as users tend to receive news from sources that are close to their opinions \cite{allcott2017social,guess2018selective}. 

\paragraph{Game Theory}
Game theory provides a mathematical framework for studying interactions in a competitive environment\cite{fudenberg1991game}. 
Recent advances in sequential decision making and reinforcement learning enable research to solve complex Markov games with many players\cite{yang2020overview,vo2018rise,grau2018balancing}. These advancements opened a new venue to study decision-making in realistic scenarios such as social interactions. Recent research presents a mathematical model for the motivation behind users sharing misinformation on social media, analyzing through multi-player Bayesian game and extensively studying the resulting equilibrium for possible solutions\cite{kurvers2021strategic, acemoglu2023model,aridor2024economics}.

Recent literature employs evolutionary game theory to enhance strategic responses by governing bodies aiming to debunk misinformation during crises such as COVID-19 \cite{vemprala2021debunking}. Additional studies model scenarios where players' misconceptions about game rules, termed misinformation games, profoundly affect outcomes, enriching our understanding of strategic misinformation \cite{varsos2021study}. Furthermore, approaches like prebunking are proposed to fortify resistance against misinformation, which involves exposing individuals to diluted forms of misleading strategies to boost media literacy and resilience against fake news \cite{roozenbeek2019fake}.

Despite these investigations addressing both online misinformation and the cognitive processes influencing individuals' propensity to propagate false information, the impact that competing news outlets have on the public's exposure to misinformation remains inadequately explored. Our objective is to address this deficiency by presenting a comprehensive mathematical framework that delineates the competitive interactions among news sources and scrutinizes how misinformation influences these competitive dynamics.

\paragraph{Opinion Dynamics}

Traditional opinion dynamics models offer insights into various social network behaviors, from polarization to consensus. The voter model \cite{holley1975ergodic}, for example, simulates opinion adoption through random interactions. The DeGroot model describes agents updating their beliefs based on the weighted averages of their neighbors' opinions \cite{degroot1974reaching}. The Friedkin-Johnson opinion model and the bounded confidence model are among the most notable derivatives of the DeGroot model. The Friedkin-Johnson model \cite{friedkin1997social} considers how individual biases influence interactions within networks, and the bounded confidence model \cite{hegselmann2002opinion} limits interactions to those with similar opinions. Although these models emphasize how microinteractions on social networks shape public opinion, larger forces such as influencers, political elites, and media corporations play a predominant role in molding public opinion \cite{helfmann2023modelling}.

We distinctly categorize news sources from the general public within the framework of a multilayered architecture for opinion dynamic evolution \cite{helfmann2023modelling}.  Further, we integrate an innovative mechanism to simulate the increased attention that news sources gain, albeit at the expense of their credibility, when they disseminate misinformation. By factoring in the susceptibility of users, our model elucidates the extent of influence that news sources exert on shaping public opinion. Subsequently, we illustrate that by integrating this alteration into established opinion dynamic models and resolving the equilibrium concerning the competition among news outlets, one can elucidate multiple characteristics associated with the development of social networks, such as ideological polarization and misinformation exposure distribution.

Research in other areas also investigates opinion dynamics and learning within social networks, focusing on how belief systems evolve through Bayesian and non-Bayesian mechanisms and how network structures influence consensus formation and misinformation propagation\cite{acemoglu2010spread,acemoglu2011opinion}. Recent studies delve into the complex interactions between media outlets and their impact on social consensus\cite{lang2022opinion}. One investigation introduces a model in which media competition and audience-oriented dynamics coexist. Various communication strategies can lead to stable pluralities or fragmented public opinions\cite{quattrociocchi2014opinion}. Additionally, some research approaches the competition for public influence as a zero-sum game, assessing equilibrium outcomes based on the information available to participants\cite{mandel2020dynamic}.


\textit{This paper} introduces a mathematical model to explain the mechanism by which competitive dynamics among various news sources can incentivize the spread of misinformation. In contrast to evaluating a singular event of misinformation propagation through social media platforms, we focus on the sequential nature of decision-making processes by news sources and their prolonged impact over time. Through detailed analysis of the equilibrium, we decode the logical incentives driving news sources toward disseminating misinformation. Subsequently, drawing from the nuanced insights gained through our analysis, we propose strategic interventions to shift the equilibria towards cultivating a reliable and trustworthy information ecosystem.

\section*{Opinion Dynamics in Presence of Misinformation}


There are several theoretical frameworks that describe the dissemination of fake news and rumors on social media platforms\cite{nguyen2012containment,de2013anatomy}. Although these frameworks estimate the diffusion of a single piece of misinformation---for instance, the dynamics of scientific rumors within social networks \cite{de2013anatomy}---they do not address impact of ongoing and recurrent exposure to false information over individual opinions. Research indicates that recurrent exposure to misinformation can foster an illusory truth effect \cite{hassan2021effects,fazio2015knowledge}, which subsequently may facilitate the propagation of misinformation\cite{vellani2023illusory}. In this paper, the focal point is not the credence individuals have in misinformation, but rather the methods employed by news sources to utilize misinformation to win over public attention.

To understand the influence of misinformation on public opinion, we assume that news sources can share misinformation to boost engagement, gaining additional influence over public opinion at the expense of damaging their credibility and diminishing their influence over skeptical individuals over time \cite{allcott2017social}. 

We employ three main concepts to model this phenomenon: source credibility, individual susceptibility, and persuasive appeal to misinformation.
The rapid expansion of social media has transformed communication dynamics and the way individuals consume news, challenging the monopoly traditional media once held over the information ecosystem. In response, we propose a model that accounts for both individual-to-individual and source-to-individual interactions within a population. Figures \ref{fig:Ad-Font} (c-f) show the key interactions in the proposed model and demonstrate the differing influence of sources with different credibility spreading factual or false information. Note that the influence of news on individuals differs based on the source credibility and news content, as well as individuals' susceptibilities. We emphasize that our definition of news source encompasses traditional media, elites, and influential groups that frequently disseminate information.


We consider a population of $N$ individuals, each holding opinion $x^i_t \in \mathcal{X} \subset \mathbb{R}$,  interacting amongst themselves and with $M$ news sources. News Sources possess fixed opinions $y^m$ for all $m \in [M]$. For integer $P$, we define the set of all positive integers up to $P$ by $[\,P\,]:=\{1,\,\cdots\,,P\}$. We denote the opinion of all $N$ individuals and all $M$ sources by vector ${ x}_t := [\,x^1_t\,,\,\cdots\,,\,x^N_t\,] \in \mathbb{R}^N$, ${ y}_t :=[\,y^1_t\,,\,\cdots\,,\, y^M_t\,]\in \mathbb{R}^M$ respectively.

\paragraph{Source Credibility} We assume that the opinions of a news source remain static, and the evolution of individual opinions does not influence it. Sources choose to disseminate factual news or misinformation at each time step. While distributing misinformation can attract a broader audience and potentially gain more influence over individuals due to its sensational nature, it also carries a risk. The public may perceive the news sources that frequently share misinformation as unreliable. This perception of untrustworthiness diminishes their impact, particularly affecting their influence on a skeptical audience.

We define the credibility of a source based on the frequency with which it disseminates misinformation. This definition is consistent with credibility assessment methodologies used by media evaluator companies, such as Ad Fontes Media and the Media-Bias-Fact-Check website. These organizations assess news source credibility by randomly sampling news distributed by each source and evaluating the reliability of the sample information.
In the proposed model, source $m$, at time step $t$, take action $a^m_t \in \{0~,~1\}$, where $ a^m_t = 0$ indicates disseminating misinformation and $ a^m_t = 1$ indicates distributing factual news. We denote the credibility for source $m$, by $c^m_t$, which evolves by the following convex combination rule,
\begin{equation}\label{eqn.cred_dyn}
    c^m_{t+1} = \lambda c^m_t + (1-\lambda) a_t^m,
\end{equation}
where $\lambda \in (0~,~1)$ represent the memory parameter for the community. When $\lambda$ is close to $1$, the public has a long memory. In this case, the public continues to associate news sources with low credibility for past misinformation, even if their current content is factual. Conversely, when $\lambda$ takes values close to zero, it suggests that the public tends to forget the actions of news sources quickly. Equation \ref{eqn.cred_dyn} illustrates how disseminating misinformation diminishes credibility while sharing factual news slightly enhances it.  We assume that all news sources are initially deemed credible, with an initial credibility score of $c^m_0=1$  for every source $m \in [M]$.

\paragraph{Individual Susceptibility} Not all individuals reflect on source credibility equally. For some, the credibility of a news source significantly impacts their acceptance or rejection of the content provided. Others may find the credibility of a source irrelevant to their judgment \cite{roozenbeek2020susceptibility}. To model individual reflection on the news from sources with different credibility, we employ susceptibility, denoted by  $s^i \in [0~,~1]$, for each individual $i \in [N]$. Susceptibility quantifies an individual vulnerability to accepting information from low-credibility sources\cite{van2022misinformation}. The higher susceptibility, the more likely an individual will be influenced by information from less credible sources. For instance, an individual with a susceptibility score of $1$ would regard the credibility of a news source as completely irrelevant, accepting information regardless of the source credibility. Conversely, individuals resilient to misinformation are more discerning, tending to trust and accept information primarily from sources that are deemed credible. This modeling approach allows us to simulate a realistic spectrum of responses to news. In this paper, we set the individual susceptibility based on beta distribution, i.e., $s^i\sim \mathfrak{B}(\beta_1,\beta_2)$ for all $i \in [N]$, where $\beta_1,~ \beta_2\geq 1$ are shape parameters.
When $\beta_1/\beta_2=1$ resulting probability density function becomes symmetric. $\beta_1/\beta_2 \rightarrow \infty$ or $\beta_1/\beta_2 \rightarrow 0$ indicates that the average shifting toward $1$ or $0$ respectively.

The definition of susceptibility we use in this paper diverges somewhat from the existing literature. In our model, individuals are influenced by news solely based on the credibility of their sources, instead of their contents. This differs from reality, where individuals factor in both the source's credibility and the content of the news\cite{metzger2010social}. For instance, skeptical individuals tend to validate each piece of news independently; however, even this group is more likely to accept misinformation from a credible source compared to an uncredible one\cite{horne2017just}. This assumption aims to capture the broad effects of individual trust and reflect real-world news consumption\cite{tsfati2003people}.  While our model simplifies susceptibility, it broadly aligns with public behavior by focusing on the trust in sources rather than the evaluation of specific news content. Future extensions of this paper could explore the detailed effects of how different types of individuals consume and accept news, as individuals likely fall into one of four categories based on susceptibility to true or false information \cite{lewandowsky2017beyond}: (1) the cynical or obstinate audience (low susceptibility to both), (2) the gullible audience (high susceptibility to both), (3) the biased or deluded audience (low susceptibility to true, high susceptibility to false), and (4) the discerning audience (low susceptibility to false, high susceptibility to true). Each group becomes either uninformed, confused, misinformed, or well-informed, but our model focuses specifically on how source credibility drives these outcomes, without delving into content evaluation.

\paragraph{Opinion Evolution} We model the overall force that influences the opinion of individual $i$ by the result of cumulative interaction forces between other individuals and $i$ (Social influence) combined with the forces of news source on $i$ (Media influence).
We model the evolution of opinion through a discrete stochastic process given by 
\begin{equation}\label{eqn.opn_dyn}
    x^i_{t+1} = \underbrace{\dfrac{a}{A^i_t} \sum_{j=1}^N \phi(\lvert x^i_t-x^j_t\rvert) (x^j_t-x^i_t)}_{\textrm{Social influence}} + \underbrace{\dfrac{b}{B^i_t} \sum_{m=1}^M \psi(\lvert x^i_t-y^m\rvert,c^m_t,a^m_t,s^i) (y^m-x^i_t)}_{\textrm{Media influence}}  + \sigma w^i_t,
\end{equation}
where $w^i_t$ is scalar i.i.d random variable drawn from normal distribution, $\mathcal{N}(0,1)$, and $\sigma>0$ denotes the strength of disturbance.
The disturbance models unknown interactions within or outside the community that affect an individual's opinion.
Non-negative scalar functions $\phi$, $\psi$ model interaction forces for social and media influence, respectively. We normalize the interaction forces to ensure a unitary effect over all individuals by $A^i_t:=\sum_{j=1}^N \phi(\lvert x^i_t-x^j_t\rvert)$ and $B^i_t:=\sum_{m=1}^M \psi(\lvert x^i_t-y^m\rvert,c^m_t,a^m_t,s^i)$. The parameters $a,~b>0$ model the amplitude of media and social influence on individuals, respectively.
Various choices of $\phi$ could recover mainstream models, such as DeGroot model\cite{degroot1974reaching} for $\phi(r) = 1$, and bounded confidence\cite{hegselmann2002opinion} for $\phi(r) = {  1}_{[0,d]}(r)$. In this paper, we consider the exponential decay interaction\cite{mas2010individualization}, i.e., $\phi(r)= \exp({-\kappa r})$ for some $\kappa>0$. 
We specifically select the exponential decaying interaction function to effectively model homophily—the tendency of individuals with similar ideologies to associate with and influence each other more significantly than those with differing views. 
Therefore, in this model, two individuals with the same ideological stance exert a notably higher influence on each other's opinions than those with divergent ideologies. This approach reflects the natural propensity for like-minded individuals to reinforce each other's beliefs, amplifying the homophilic effect on the evolution of opinions.

\paragraph{Misinformation Gain}
Modeling the interactions between news sources and individuals presents a more complex challenge compared to individual-individual interaction, particularly when considering the variable effects of source credibility on individuals with different susceptibility levels. Disseminating misinformation may attract a larger audience due to its sensational and controversial nature\cite{vosoughi2018spread,graves2019information}, but it compromises source credibility. The credibility of a news source significantly influences its impact, particularly among skeptical individuals. This creates a dilemma for news sources: gain additional influence or risk damaging their reputation.

To this end, we model the influence of news sources based on the interplay between the source credibility, individual susceptibility, and the actions news source takes by
\begin{equation}
    \psi(r,c,a,s) = \exp[{-\hat\kappa f(a)g(c,s)r}],
\end{equation}
where $f(a)=1+\eta a$ for some \textit{misinformation gain} $\eta>0$, and $g(c,s)=1+\zeta(1-c)(1-s)$ for some \textit{credibility gain} $\zeta>0$. The function $f$ models the boost in immediate influence a media has on the public when it disseminates misinformation and sensational news aimed to attract a broad audience. 
Misinformation gain $\eta$ determines the additional influence a source gains when sharing misinformation.

People tend to accept news from sources that share similar opinions \cite{mccombs2011news,feldman2011opinion}. We model this effect by letting the media influence multiplier $\psi$ decay exponentially with the distance $r$ between the opinions of the source and the individual. This model ensures that individuals are most influenced by news sources whose opinions align closely with their own. $\hat{\kappa}>0$ indicates the level of influence news sources have in general over the public. In general, news sources have access to a broader audience, and thus, we can assume $\hat \kappa \leq \kappa\ $.

Conversely, we design the function $g$ to account for the influence of news source credibility on individuals. 
Figure \ref{fig:Ad-Font}(b) illustrates the media influence intensity with respect to the susceptibility of individuals and the credibility of a news source.  The susceptibility of individuals becomes irrelevant when news originates from a highly credible source ($c=1$). However, information distributed by low-credible sources has less influence on skeptical individuals. Credibility gain $\xi$,  reflects the penalty applied to sources with low credibility. This design underscores how the interaction between news source credibility and individual susceptibility shapes the overall effect of media influence. 

\paragraph{Misinformation exposure} 
Since our model considers the possibility of news sources disseminating misinformation, we can assess individual exposure to misinformation, effectively estimating their reliance on misleading and sensational narratives \cite{mosleh2022measuring}.
Misinformation exposure measures the probability that an individual is exposed to misinformation shared by sources over time. To this end, we define the misinformation exposure score for each individual, $i$, during $T$ duration by
\begin{equation}
    \Gamma_i=\dfrac{1}{TM}\sum_{t=0}^T\sum_{m=1}^M e^{-\hat \kappa \lvert x^i_t-y^m\rvert} (1-a^m_t).
\end{equation}
where $e^{-\hat \kappa \lvert x^i_t-y^m\rvert}$ measures the chance that individual $i$ hear the an story shared by source $m$. 

Through this paper, we assume that the network includes $N=500$ individuals and $M=10$ sources unless otherwise stated. The initial distribution is randomly selected from the distribution $\rho_0$, where $\rho_0$ is assumed to be uniform with support $\mathcal{X}$. We let  $\eta =1$ and $\xi=2$ represent the attention gained through sharing misinformation(misinformation gain) and the attention lost for being perceived as untrustworthy (credibility gain), respectively. The homophily coefficient, for the user-user, is set by $\kappa =20$, and for the user-media by $\hat \kappa =5$, assuming news sources have a stronger homophilic influence. We let $a=b=h$, where $h=0.1$ and noise amplitude is $\sigma=0.1\sqrt{h}$. 
The susceptibility is distributed by a beta distribution with shape parameters $\beta_1=3$, and $\beta_2=2$.

\begin{figure}
    \centering
    \includegraphics[width=\textwidth,trim={00 00 00 0},clip]{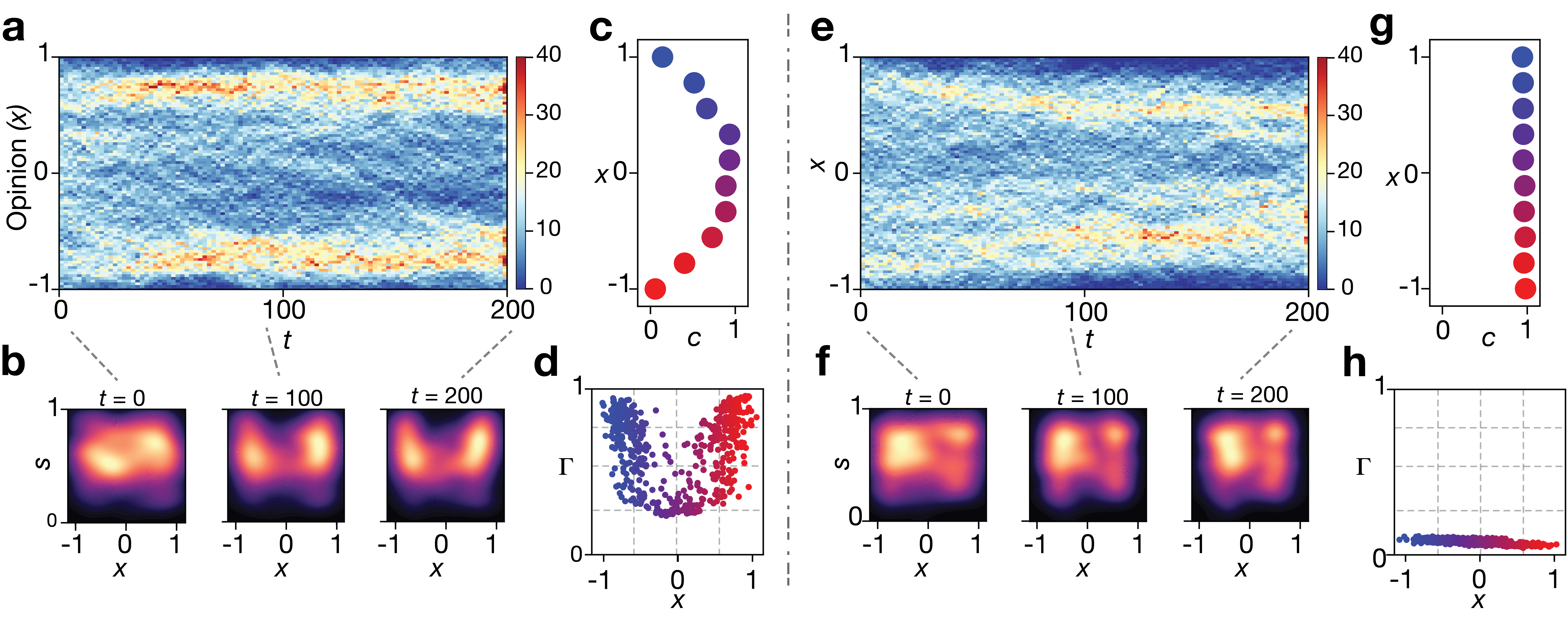}
    \caption{ Opinion dynamics become polarized when news source strategies for disseminating misinformation reflect real-world policies. \textbf{a,e.} Evolution of opinions among $N=500$ individuals for two different strategies \textbf{c,g.} with \linebreak[4] $\eta=1,~\xi=2,~\beta_1=3,~\beta_2=2$. \textbf{b,f.} Snapshots of susceptibility and opinion distribution over time steps $t=\{0,100,200\}$. 
     Initially, hyper-partisan sources start with high credibility while frequently disseminating misinformation, thereby gaining significant influence over the public. However, as their integrity wanes, individuals resilient to misinformation shift towards the center. \textbf{d,h.} Final misinformation exposure, $\Gamma$, with respect to opinion bias. Unbiased individuals are exposed to considerably less misinformation in comparison to others.}
    \label{fig:OP_Dyn}
\end{figure}

\subsection*{Information Structure}

After a news source shares information, fact-checking based on available information becomes feasible. We assume that information about the credibility of news sources are public and accessible to all stakeholders. We define the state as the information about all individuals' opinions and sources credibility, represented by ${q}_t = [{x}_t; {c}_t]$. Realistically, estimating every opinion is infeasible; thus, we assume that news sources can access only the discretized distribution of opinions. This assumption stems from the fact that capturing public opinion typically relies on time-consuming and expensive methods such as surveys and polls, which only estimate opinion distributions. We divide the opinion space into $l$ equal intervals and report the number of users in each interval divided by the population size by $z_t \in \mathbb{R}^l$. Each source can observe $o_t = [z_t;c_t]$ and decide to disseminate misinformation through strategy $\pi^m(o_t)$, by choosing an action $a^m_t \sim \pi^m(o_t)$.

Figure \ref{fig:Ad-Font}(a) presents the credibility and opinion biases for $223$ news sources. It illustrates a trend where centrally positioned sources are deemed most credible, while radical sources are more likely to disseminate misinformation. If news sources adhere to a fixed policy, they will converge to stationary states. In this case, the credibility score represents the probability of sharing factual news, i.e., $c^m_\infty = \mathbb{P}(a^m=1)$. Thus, Figure \ref{fig:Ad-Font}(a) estimates policies and credibility, assuming sources are taking fixed actions.

Figure \ref{fig:OP_Dyn} shows the distribution of opinions for $t=200$ steps under a consistent policy, $\pi^m(o_t) = \pi^m$ for all $m \in [M]$, across two scenarios. First, news sources strictly limit misinformation, only sharing factual content with occasional errors (Figure \ref{fig:OP_Dyn} e-h). In the second scenario, hyper-partisan sources disseminate significantly more misinformation compared to centrist sources (Figure \ref{fig:OP_Dyn} a-d). When the sources refrain from sharing misinformation, public opinion reaches a consensus. When sources abstain from disseminating misinformation, public opinion converges towards consensus, reflecting interactions free of misinformation.

\begin{wrapfigure}{r}{0.4\textwidth}
\vspace{-20pt}
  \begin{center}
    \includegraphics[width=0.4\textwidth]{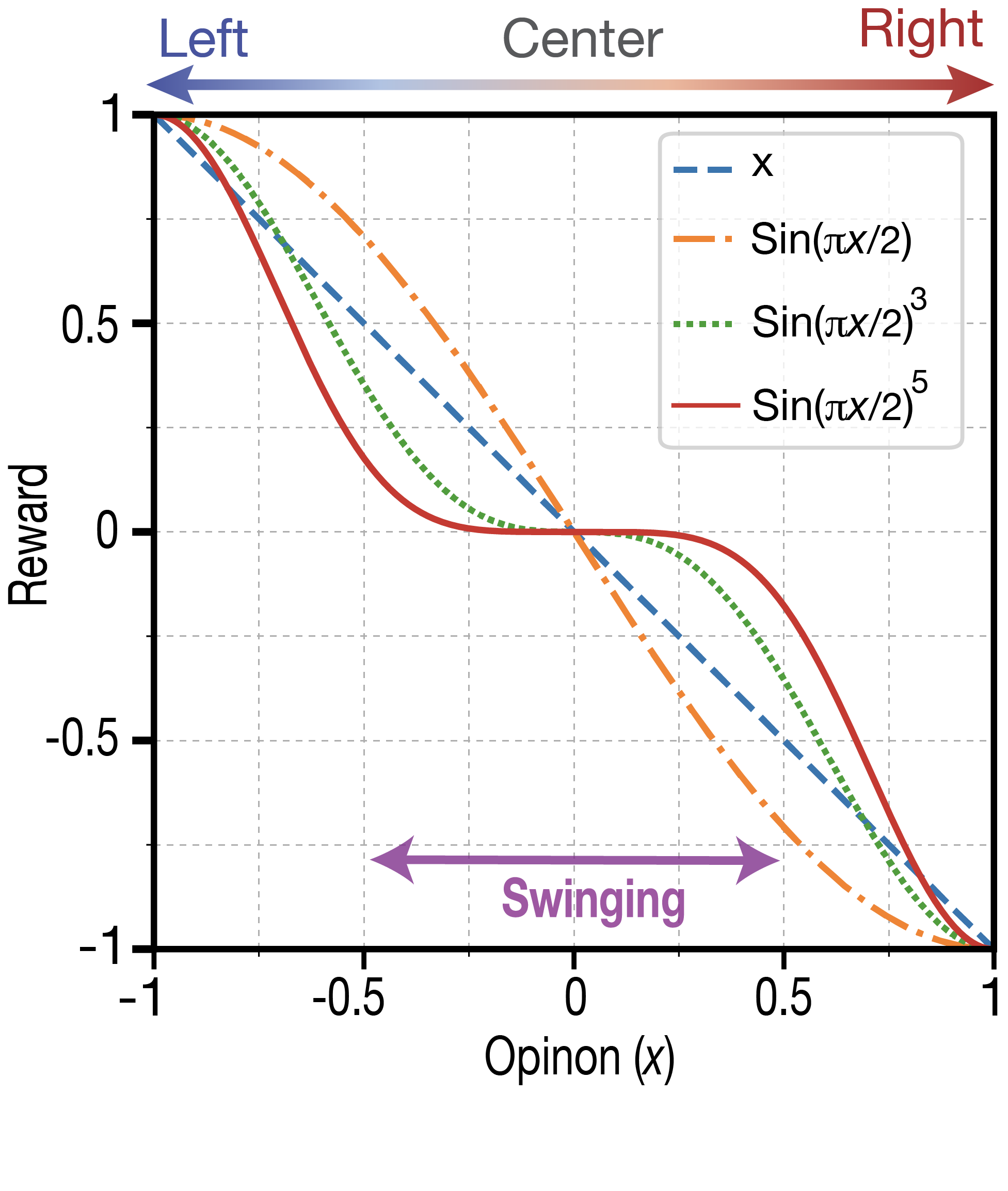}
    \caption{Different candidates for the reward function $r(x)$. The players seek to radicalize public opinion in their favor and minimize the influence of the opponent. The exponent of the reward function controls players' preference for radicalization. Larger exponents (red, green) indicate a stronger preference for radicalized public opinion.}\label{fig:reward}
    \vspace{-10pt}
  \end{center}
\end{wrapfigure}

Public opinion becomes polarized when news sources employ the strategy reported by Ad-Font media, as shown in Figure \ref{fig:Ad-Font}(a). Initially, all sources begin with equal credibility, $c_0^m=1$ for all $m \in [M]$. At this stage, all sources are relatively credible regardless of their actions, meaning that disseminating misinformation affects all users, irrespective of their susceptibility. This scenario creates an opportunity for dishonest news sources to attract both susceptible and skeptical individuals. As a result, radical sources quickly established echo chambers that included users of all types. Over time, as the credibility of these radical sources wanes, skeptical individuals move toward the center, while susceptible users stay within the echo chambers. If skeptical individuals generate sufficient momentum, this shift can depolarize the community. However, as illustrated in Figure \ref{fig:OP_Dyn}(a), such migration, even with the influence of centered sources, often fails to prevent stable polarization. The proposed model's results emphasize misinformation's role in forming and stabilizing echo chambers, especially under diverse and non-uniform strategies that reflect real-world dynamics.

In Figure \ref{fig:OP_Dyn}(d,h), we analyze the final misinformation exposure of the population. When news sources prioritize factual reporting, overall misinformation exposure remains notably low. Conversely, when source policies reflect real-world scenarios, individuals with radical opinions are significantly exposed to misinformation. These results are consistent with studies on misinformation exposure on platforms like Twitter \cite{mosleh2022measuring}, demonstrating that the proposed model effectively captures various aspects of misinformation dissemination. Although the results highlight several phenomena, such as the formation of echo chambers in the presence of misinformation and misinformation exposure distribution for individuals, they do not explain why news sources adopt strategies like those depicted in Figure \ref{fig:Ad-Font}. Understanding the competitive dynamics among news sources that foster misinformation dissemination is crucial for designing policies to depolarize public opinion and reduce misinformation exposure. We delve deeper into the decision-making processes of news sources, exploring how competition for public influence can lead to the spread of misinformation. This investigation will aid in developing more informed and effective media regulations and public policies.

\section*{Equilibrium Concept}

In the previous section, we discussed the interplay between news sources disseminating misinformation and public opinion evolution. To understand the decision-making processes of news sources in sharing misinformation, we extend this analysis to investigate their strategic behaviors as they compete to increase their influence on public opinion \cite{keppo2022learning, kurvers2021strategic}. We conceptualize this competition as a zero-sum game, where two players, ${L~,~R}$, each controlling half of the sources, vie for public opinion. Specifically, we consider sources on either side of the political spectrum as coordinated teams seeking to increase their respective parties' influence.

Player $L$ controls the news sources with $y^m <0$, i.e., $m \in \mathcal{M}_L=\{1, \cdots, M/2\}$ while player $R$ controls sources with $y^m<0$, i.e., $ m \in \mathcal{M}_R= \{M/2, \cdots, M\}$. We denote the ${a}^R_t \in \mathcal{A}_R $ and ${  a}^L_t \in \mathcal{A}_L$ as the vector of actions taken by player $R$ and $L$ respectively. Each player has $2^{\frac{M}{2}}$ possible actions to choose from and takes actions based on a policy ${ \pi}^\star ({  a}^\star|{  q}_t)$ where $\star \in \{R~,~L\}$.
The players want to shift public opinion in their favor by maximizing the population with similar opinions and minimizing those with opposing opinions. We formulate the competition as a zero-sum game, given by

\begin{equation}\label{eqn.game}
    \begin{aligned}
        \max_{{ \pi}^L} \min_{{ \pi}^R}  \quad &  J:= \mathbb{E} \Big[ \sum_{k=1}^\infty
        ~ \gamma^{k}\, r({x}_{k}) \Big],\\
        \textrm{s.t.} \quad & \textrm{\eqref{eqn.opn_dyn}}, ~ \forall~ i \in [N],\\
          & \textrm{\eqref{eqn.cred_dyn}},~ \forall~ m \in [M],  \\
          & x_0 \sim \rho_0, ~~c_0 = {\bf 1}, \\
    \end{aligned}
\end{equation}

where player $L$ wants to maximize, and player $R$ wants to minimize the cost function $J$ and $\gamma$ is the discount factor. We draw the initial distribution of opinions, $x_0$, from uniform distribution $\rho_0$.
The players want to pull the public opinion in their direction. Therefore, we define the running cost, $r(\cdot,\cdot):\mathbb{R}^{M+N}\rightarrow \mathbb{R}$, by

\begin{equation}
    r({x}_t,{c}_t) = -\sum_{i=1}^N sin(\varpi x_t^i)^ \vartheta.
\end{equation}

Parameters $\varpi := \frac{\pi}{2}$ and $\vartheta := 5$ represent the opinion strength and steepness target where players aim to shift public opinion. Increasing $\varphi$ shifts the maximum of the reward function toward the center, with the reward peaking at $x=1$ when $\varpi = \frac{\pi}{2}$. Meanwhile, $\vartheta$ controls the rate of change in the reward function; increasing it leads to a rapid and steep rise or decline in rewards.
Figure \ref{fig:reward} illustrates various candidates for the reward function. Since the objective of each player is to radicalize the community, individuals who are likely to swing between parties are undesirable as they can shift allegiances effortlessly. To address this aspect, we set $\vartheta:= 5$. This adjustment ensures that the rewards are considerably smaller for individuals in the swinging region and higher for those more radically aligned. This approach incentivizes players to focus their efforts on solidifying and expanding their base of radical supporters, thereby maximizing their influence within these segments. In addition, changing $\varpi$  allows us to tune the level of radicalized players aim for. Note that the cost function outlined in Equation \eqref{eqn.game} does not impose direct penalties on disseminating fake news or the credibility of news sources. Instead, it indirectly penalizes misinformation dissemination over time through credibility erosion, which manifests as a diminished influence on individuals with low susceptibility. 

To measure the polarization of opinion distribution, we employ the bimodality coefficient\cite{pfister2013good}, $\varsigma$, defined by 
\begin{equation}
    \varsigma = \frac{\chi_3^2+1}{\chi_4},
\end{equation}
where $\chi_3$ and $\chi_4$ represent the skewness and kurtosis for the opinion distribution. A bimodality coefficient smaller than $\textrm{BT}:= \frac {5}{9}$ suggests that the distribution is likely unimodal, indicating less polarization. Conversely, a coefficient greater than this threshold indicates a multimodal distribution, suggesting significant polarization. This measure assists us in evaluating the effectiveness of sources strategies in influencing public opinion, particularly in terms of polarization.

\subsection*{Quantal Response Equilibrium}

Human decision-making often deviates from complete rationality\cite{gigerenzer2001rethinking,evans2013rationality}. Similarly, media corporations and influential groups run by humans can make irrational and sometimes random decisions. Recognizing that the Nash equilibrium fails to capture bounded rationality, we adopt the quantal response equilibrium to model the probabilistic nature of human decision-making \cite{mckelvey1995quantal,mckelvey1998quantal}, which provides a generalization of Nash equilibria to scenarios where the players necessarily have some randomness in their strategies.
For the rest of the paper, we denote the quantal response equilibrium by equilibrium unless stated otherwise.

For two given policies $\mu$, and $\mu'$, we define the  Kullback-Leibler divergence from $\mu$ to $\mu'$ by
\[
KL(\,\mu\,||\,\mu'\,)=\sum_a \,\mu(a) \, \log(  \, \frac{ \, \mu(a) \, }{ \, \mu'(a) \, } \, ).
\]
To model bounded rationality, we assume each player's preferred policy has a bounded Kullback-Leibler divergence from a reference policy $\bar {  \pi}$.
We add rationality constraints to the game \eqref{eqn.game} to model the competition with bounded rationality.
\linebreak[4] A prominent technique to solve the game with rationality constraint is to employ Lagrange multipliers, i.e., integrating the constraint into the objective function and transforming the original problem into an unconstrained equivalent \cite{grau2018balancing}. We then seek to find the value function through
\begin{equation} \label{eqn.valu-func}
    V^* = \max_{{ \pi}^L} \min_{{ \pi}^R}   \mathbb{E}\Big[ \sum_{k=1}^{\infty} ~ \gamma^{k}\, \big ( R(q_t) 
        -\frac{1}{\tau_R} \log(\frac{{ \pi}^R ({ a}^R_t|o_t)}{\bar{ \pi}(a_t)}) 
        +\frac{1}{\tau_L} \log(\frac{{ \pi}^L ({ a}^L_t|o_t)}{\bar{ \pi}(a_t)}) \big )\Big],
\end{equation}
where $\tau_R$, and $\tau_L$  denote the rationality of the players. Due to the high dimensionality of this game, solving it with optimal control techniques is impractical. Therefore, we employ function approximation to solve for the Value function, achieving a local minimum that approximates the optimal policies for the game.

While our initial model assumes complete state information observability, tracking each individual's opinion in practice is challenging and raises ethical concerns. Therefore, in subsequent sections, we assume that players have access to only partial observations. This adjustment acknowledges the practical limitations and ethical considerations in monitoring individual opinions. Although this change might cause the equilibrium to deviate from what is achievable under complete observability, it leads to more realistic modeling.

\begin{figure}
    \centering
    \includegraphics[width=1 \linewidth]{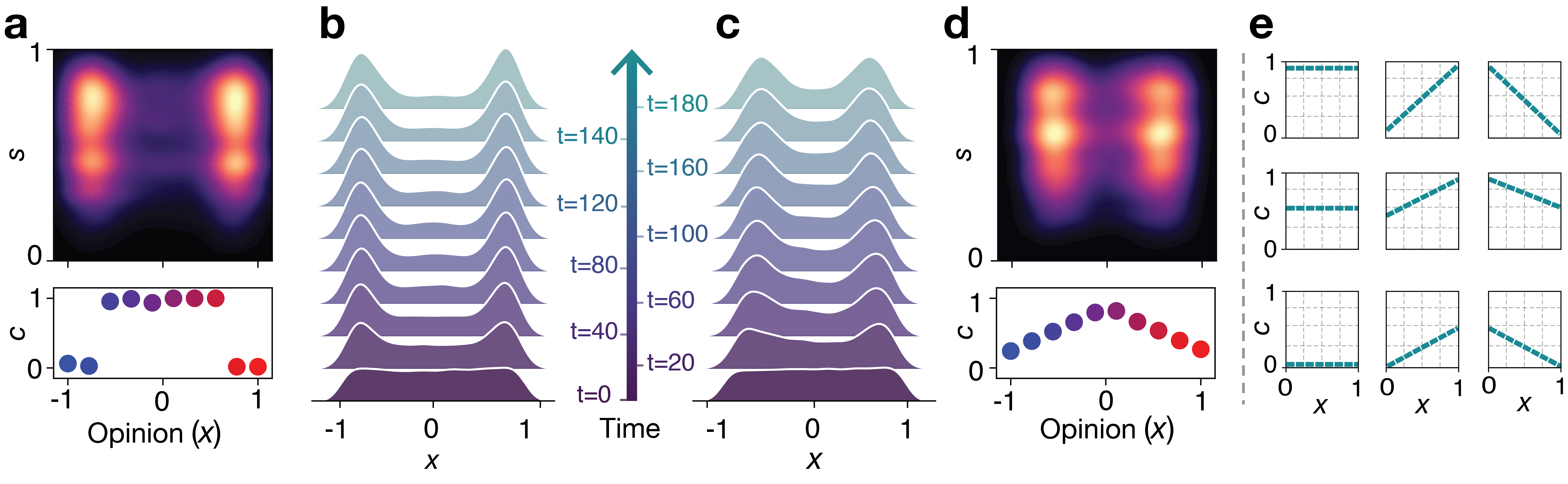}
    \caption{  The equilibrium for $M^2$ actions \textbf{a,b.} and nine selected actions \textbf{c,d.}. Both cases entirely polarize public opinion, but the complete action achieves greater separation between peaks. \textbf{(b)} The evolution of opinion when each source can choose any action at each time for the equilibrium. \textbf{c} The evolution of opinion distributions when news source actions are limited to strategies depicted in \textbf{e}. \textbf{a,d.} The final susceptibility-opinion distribution for complete action and limited strategies, respectively. 
     \textbf{e.} The 9  action profiles that players can choose from in each round.}
    \label{fig:Complete-Action}
\end{figure}

\paragraph{Limited Action}  
Assuming that news sources immediately respond to each observation is impractical. Public opinion is difficult to estimate, leading to infrequent polling. More realistically, each party periodically sets and adjusts a general policy direction. Additionally, obtaining equilibria for sources that immediately respond to each observation poses scalability challenges as the action space grows exponentially with respect to the number of news sources.

To address the curse of dimensionality, we introduce a finite set of profile strategies, as illustrated in Figure \ref{fig:Complete-Action}(e). To this end, we simplify the action space by restricting players to choose from predefined strategy profiles. We then evaluate these strategies in two scenarios. First, players remain immediately responsive to each observation, but with limited action profiles to choose from at each step. Second, they consistently follow a predetermined profile throughout the competition. Thus, in the latter case, the original Markov game \eqref{eqn.game} simplifies into a matrix game, facilitating an efficient computation of the equilibrium and allowing a comprehensive sensitivity analysis.

Assume that $\mathcal{B}_L$ and $\mathcal{B}_R$ represent the set of action strategies players $L$ and $R$ can take, respectively, as shown in Figure \ref{fig:Complete-Action}-e. We define the zero-sum entropy regularized matrix game for completeness by the following
\begin{equation}\label{game.matrix}
    \max_{\mu\in \Delta(\mathcal{B}_R)} \min_{\nu\in \Delta(\mathcal{B}_L)} \mu^T A \nu + \frac{1}{\tau_R} \mathcal{H}(\mu) - \frac{1}{\tau_L} \mathcal{H}(\nu),
\end{equation}
where $A \in \mathbb{R}^{p \times p}$ stands for the payoff matrix, $\mu\in \Delta(\mathcal{B}_L)$ and $\nu \in \Delta(\mathcal{B}_R)$ represent the mixed policies for each player, defined respectively as distributions over probability simplex $\Delta(\mathcal{B}_R)$ and $\Delta(\mathcal{B}_L)$.
We use Shannon entropy, defined as $\mathcal{H} (\pi) := -\sum_i \pi_i\log (\pi_i)$,  to account for the bounded rationality of players, where $\pi$ represents an arbitrary distribution over action space. It is well known that the equilibrium for matrix games is unique and satisfies 
\begin{equation}
    \begin{cases}
        \mu^*(a) &= \dfrac{exp(\tau_L [A\nu^*]_a)}{\sum_b exp(\tau_L [A\nu^*]_b)},  \quad \textrm{for all } ~a \in \mathcal{B}_L,\\
        \,\\
        \nu^*(a) &= \dfrac{exp(\tau_R [A\mu^*]_a)}{\sum_b exp(\tau_R [A\mu^*]_b)},  \quad \textrm{for all } ~ a \in \mathcal{B}_R.
    \end{cases}
\end{equation}
The above game solutions follow as a corollary to Theorem 4 by Amos \cite{amos2019differentiable}. We use the extragradient method to find the equilibrium efficiently\cite{cen2021fast}, by estimating the payoff matrix empirically through $200$ simulations.

\section*{Results}

In this section, we discuss the numerical results. Initially, we present hyper-responsive players who exert complete control over each source's actions. Subsequently, we introduce strategies to limit the action space, demonstrating that this approach yields a reasonable approximation of the complete action space Equilibrium. We also transform the Markov game into a stochastic matrix game when players consistently select a dominant action throughout the course of the game. This transformation enables highly efficient computation of the equilibrium for matrix games. Leveraging this, we conduct a comprehensive sensitivity analysis on various factors within the proposed model.

Figure \ref{fig:Complete-Action}(a,b) shows the evolution of opinion distribution when players, having complete control over the action space, act according to the equilibrium. Although the objective is to radicalize the community, the actions lead to polarization. The results show that equilibrium forces news sources to spread misinformation extensively through hyper-partisan sources while preserving the credibility of mainstream sources. The equilibrium is exploiting the initial high credibility of radical sources to attract as many individuals as possible in the early stages and form echo chambers while maintaining high credibility for the centrist sources so it can affect the other party most efficiently.

Computing the equilibrium for a large action space is computationally demanding and presents significant challenges. To address this issue, we use a fixed set of strategies from which news sources can choose from. Figure \ref{fig:Complete-Action}(c,d) depicted the resulting equilibrium when players are limited to actions based on strategies depicted in Figure \ref{fig:Complete-Action}(e). The equilibrium for both complete and limited action spaces results in polarization, and the opinion distribution evolves similarly.

The final strategy for the complete action space in Figure \ref{fig:Complete-Action}(a) shows that for optimal policy, both players should push the credibility of news sources close to either $0$ or $1$. This result is contrary to the real-world credibility distributions of news sources given in Figure \ref{fig:Ad-Font}(a), where the credibility-opinion relation of news sources loosely follows a smooth curve. We hypothesize that this deviation from real-world behavior might be due to two effects. Firstly, real-world news sources with similar political biases often share similar news items. As such, misinformative news items diffuse between news sources that are of similar opinions. This diffusion pulls the credibility of similar news sources closer, resulting in a smoother credibility-opinion distribution. Our model does not account for such inter-news-source diffusion effects, resulting in sharp transitions in the optimal credibility-opinion distributions of both players.

\begin{figure}
    \centering
    \includegraphics[width=1\linewidth]{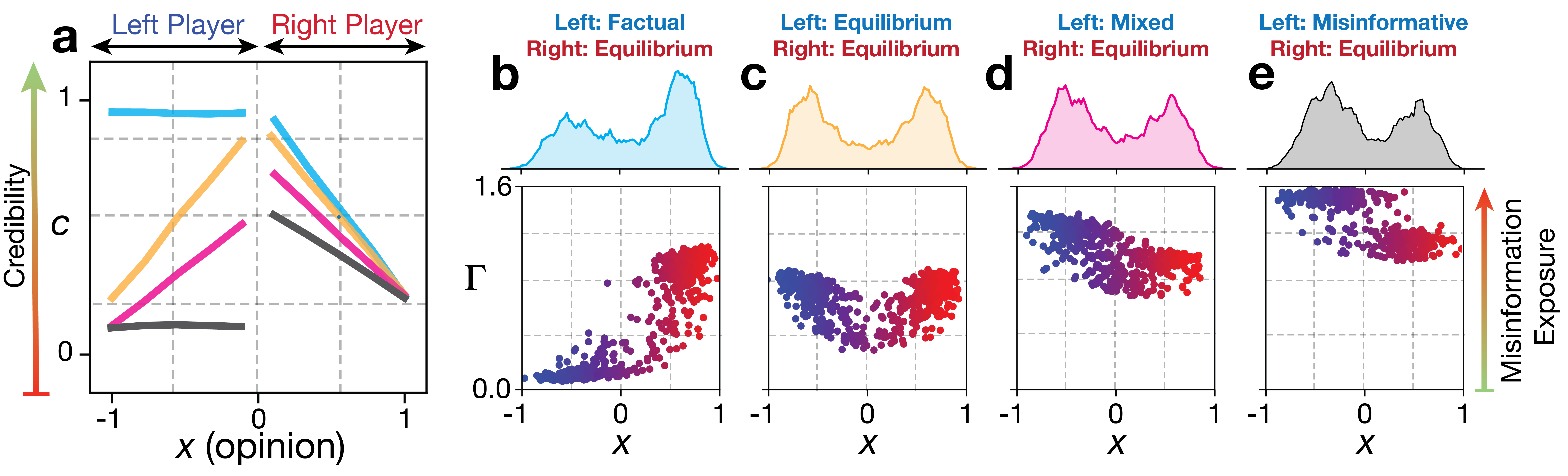}
    \caption{Whenever a player increases misinformation dissemination below equilibrium, the other player's optimal response causes it to share more misinformation. \textbf{a.} The left player changes its policy to share more misinformation by deviating from equilibrium, while the right player responds optimally. \textbf{b.} The left player avoids disseminating misinformation, allowing the right player to radicalize the community in its favor. \textbf{c.} Both players take action based on equilibrium \textbf{d.} The left player disseminates more misinformation, forcing the other player to lower its credibility accordingly.  }
    \label{fig:Mixed}
\end{figure}

Secondly, the game setup for complete action space assumes that the players have complete control over the credibility of all news sources. Such a scenario is unrealistic for a real-world setup, where political parties only have a limited amount of influence over the news coverage. Thus the results for employing a limited set of actions, shown in Figure \ref{fig:Complete-Action}(c,d), may present the game solution in a more realistic set of assumptions. However, the overall shape of the optimal credibility-opinion strategies for both complete and limited action results shows a similar trend of central news sources having higher credibility. This trend, which is also prevalent in real-world data does not seem to be sensitive to the limitations on the action space.

For the remainder of the paper, we assume that players are limited to taking actions based on the strategies presented in Figure \ref{fig:Complete-Action}(e). To better understand the decision-making processes of news sources, we explored scenarios where the left player takes sub-optimal actions. In contrast, the right player follows the optimal policy learned during training.

The results in Figure \ref{fig:Mixed} show that when one player disseminates misinformation more frequently, the other player's optimal response is to answer with increasing dissemination of misinformation.
Thus, when one player deviates from equilibrium policies and disseminates more misinformation, the community is damaged in two stages. First, it increases the exposure of misinformation among its community as a direct result of its actions. Second, the opposing player optimal response to such deviation is to reduce its credibility accordingly, further damaging both communities.  Figure \ref{fig:Mixed} illustrates this phenomenon, highlighting the reciprocal nature of misinformation dissemination. This dynamic suggests a kind of arms race in misinformation dissemination, where actions taken by one side influence and escalate the strategies of the other.

Empirical research on misinformation exposure on Twitter indicates an imbalance between parties \cite{mosleh2022measuring}. Our findings in Figure \ref{fig:Mixed} suggest that this imbalance may stem from differences in decision-making regarding misinformation dissemination across various parties.

Our results suggest that the equilibrium shown in Figure \ref{fig:Mixed} represent a stable equilibrium, meaning that starting from various initial conditions for media credibility consistently leads to the same equilibrium state. This stability is crucial, indicating that if the Left player adopts strategies based on the equilibrium it has learned, it will converge to the equilibrium depicted in Figure \ref{fig:Mixed}(c). However, it's important to note that function approximation methods, as used in our analysis, typically find local solutions. This limitation underscores the need for further research to validate the uniqueness and stability of the equilibrium.

\begin{figure}[t]
    \centering
    \includegraphics[width=1\linewidth]{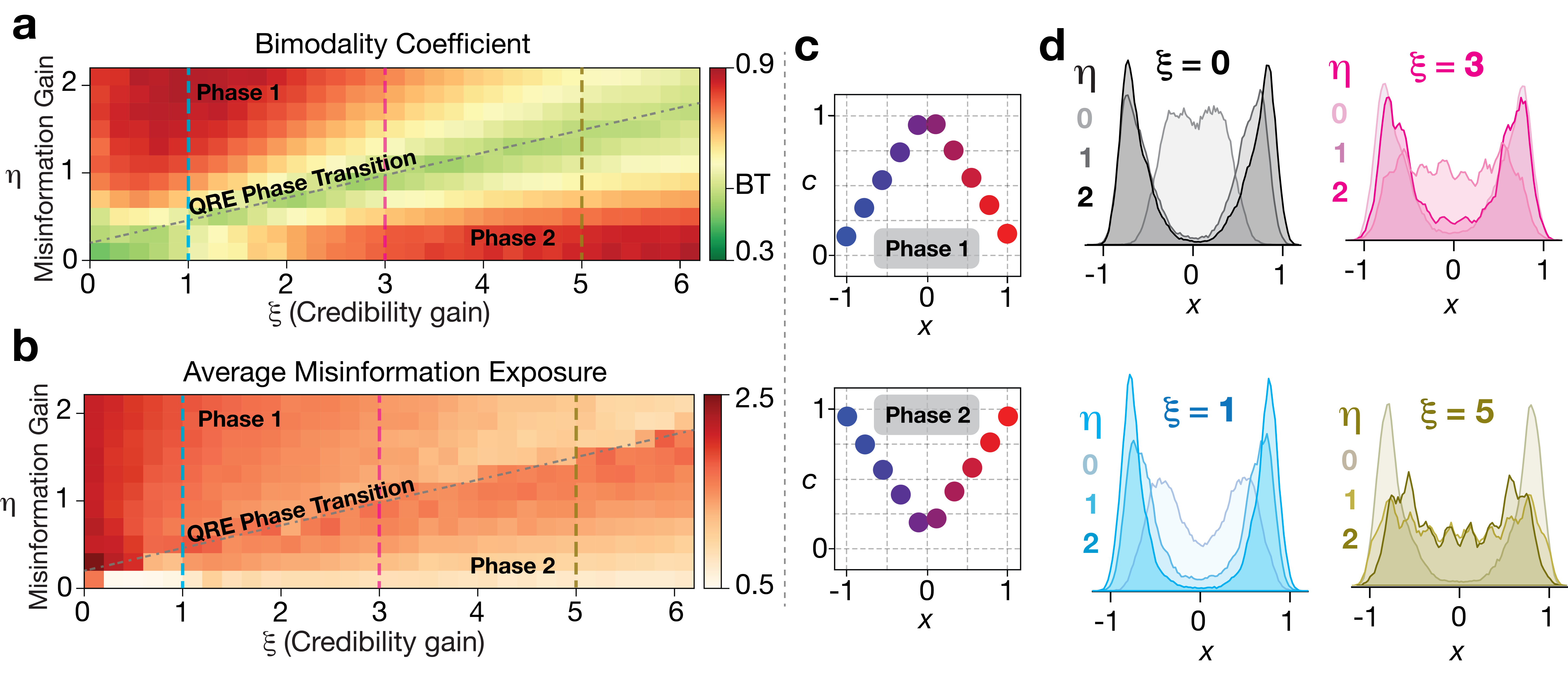}
    \caption{ Reducing attention gained from misinformation may cause a phase transition \textbf{c.} in equilibrium that depolarizes public opinion and exacerbates misinformation exposure.
    \textbf{a.} Bimodality coefficient for different values of misinformation gain, $\eta$ and credibility gain $\xi$. The green region indicates where players cannot exploit policies that lead to polarization.  \textbf{b.} When the impact of misinformation outweighs credibility, the equilibrium results in excessive misinformation exposure. \textbf{d.} Different snapshots of public opinion distributions.}
    \label{fig:xi-eta}
\end{figure}

\subsection*{Sensitivity Analysis} 
News sources often adhere to specific strategies for extended periods, especially before critical events \cite{kim2022measuring}. Consequently, it is reasonable to assume that each player consistently selects and maintains a predominant strategy throughout the game. This assumption simplifies the Markov game into a matrix game, enabling efficient computation of the equilibrium.

Leveraging this computational efficiency, we conduct an extensive sensitivity analysis focusing on key factors such as misinformation gain, credibility gain, player rationality, and the distribution of susceptible individuals. This analysis provides insight into how different choices of parameters influence the dynamics and outcomes of the game. In addition, sensitivity analysis aids in predicting how changes in the information landscape could impact public opinion and misinformation exposure. Thus, it opens new avenues for government bodies, media organizations, and societal elites to combat misinformation effectively.

\begin{figure}[t]
   \centering
   \includegraphics[width=\linewidth]{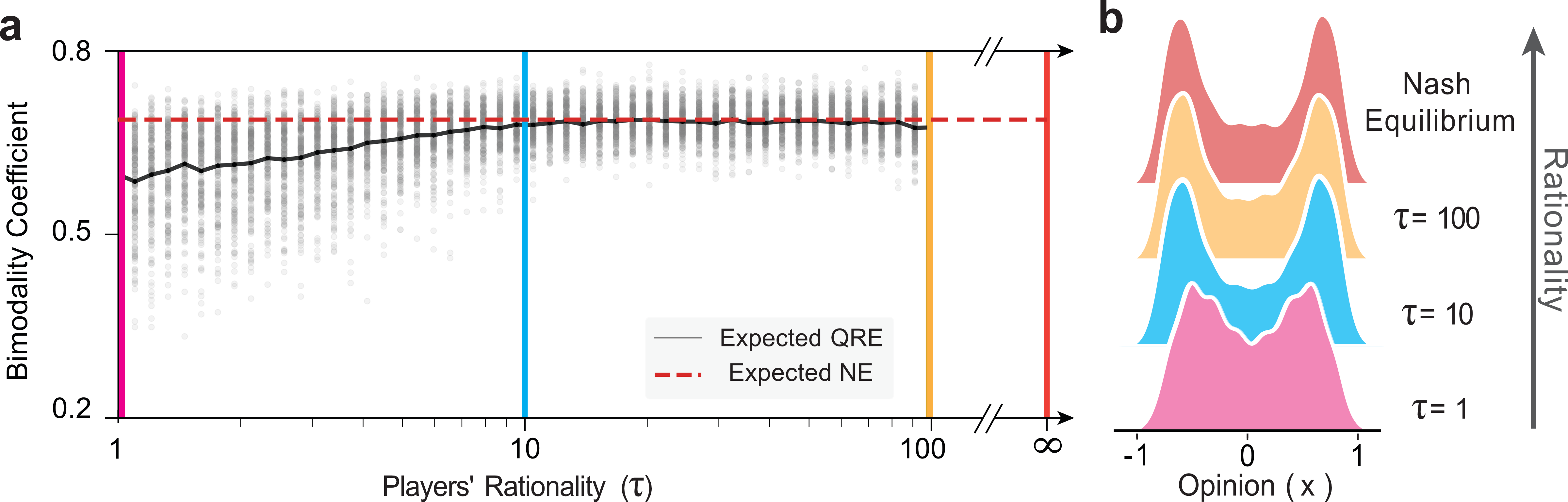}
    \caption{Increasing rationality leads to more polarized equilibrium.
    \textbf{a.} Bimodality coefficient for different levels of rationality, $\tau$.
    \textbf{b.} Slices showing opinion distributions of users at equilibrium for $\tau=\{1\,,\,10\,,\,100\}$ and Nash Equilibrium ($\tau \rightarrow \infty$).  Increasing $\tau$ makes the quantal response equilibrium computation algorithm unstable. }
    \label{fig:fixed-rationallity} 
\end{figure}

\paragraph{Misinformation and Credibility Gain}

Recall that $\xi$ assesses the influence of perceived source credibility on individuals, and $\eta$ determines the additional influence gained from disseminating misinformation. Since these parameters enforce the effects of credibility and misinformation on the community, they significantly impact the equilibrium. Figure \ref{fig:xi-eta} shows the bimodality coefficient, $\varsigma$, and the average misinformation exposure, $\bar \Gamma$ across various levels of $\eta$ (misinformation gain) and $\xi$ (credibility gain) for the equilibrium.

 When we set both $\xi$ and $\eta$  to zero, i.e.,  the actions of the news source are irrelevant, public opinion reaches a consensus. Additionally, it highlights that polarization results from the effects of misinformation on public opinion. As the misinformation gain increases, players' optimal policies shift toward more aggressive dissemination of misinformation to compete for influence, escalating polarization and raising misinformation exposure levels. 

When misinformation gain outweighs credibility gain, the optimal policies mirror the real-world credibility-opinion curve.  Increasing the credibility gain or reducing the misinformation gain leads to a phase transition in the equilibrium, as depicted in Figure \ref{fig:xi-eta}. During this transition, credibility gain dominates, making it less beneficial for players to rely on misinformation for radicalization; thus, the equilibrium policies reverse. Players then aim to enhance the credibility of their radical news source and more frequently disseminate misinformation through centrist sources. In the second phase, players target low-susceptibility individuals to create echo chambers, contrasting with the first phase, where they rely on susceptible individuals.  While this observation suggests that polarization can still occur even when credibility is deemed more influential than misinformation, the phase transition results in a less polarized community. However, phase transition significantly increases misinformation exposure. This sudden change is because when policies reverse, centrist sources disseminate misinformation more often. Thus, a depolarized community is in more danger than a polarized one in the second phase. 

Finally, our findings suggest that reducing misinformation gain is more beneficial than increasing credibility gain to reduce misinformation exposure and depolarize the community. While the results advocate for reducing misinformation gain as a more effective strategy, it is essential to acknowledge the significant practical challenges it presents. Indeed, decreasing misinformation gain requires reducing the influence of sources on the susceptible individuals, who are the primary audience of radical news sources. Therefore, for meaningful change in misinformation gain, one must enforce it from hyper-partisan sources. However, these changes would weaken the grasp of those sources on susceptible users, which makes such changes unlikely.

\paragraph{Rationality}

Adopting quantal response equilibrium allows us to evaluate the results for different levels of rationality.
The influence of player rationality on the equilibrium is another pivotal factor in the dynamics of sources' influence on public opinion. As illustrated in Figure \ref{fig:fixed-rationallity}, increasing levels of rational decision-making—approaching the Nash equilibrium as a benchmark for perfect rationality—leads players to adopt more extreme strategies. This trend results in greater polarization within the community. With heightened rationality, players more efficiently exploit hyper-partisan news sources and engage more frequently in misinformation dissemination, intensifying polarization and reinforcing their influence over the public. Notably, our results highlight that the perception of radical sources as less credible is more than just a result of irrational behavior. Instead, it reflects a calculated response to the strategic landscape in which higher rationality enhances the effectiveness of misinformation as a tool for gaining influence. 
The bimodality of the opinion distribution of the population at equilibrium remains nearly constant for $\tau > 10$. Thus, setting $\tau = 10$ provides a good approximation of the Nash equilibrium while maintaining the smoothness of the value function we define in Equation \ref{eqn.valu-func}, which is necessary for the extragradient method \cite{cen2021fast} to converge.

\begin{figure*}[t]
   \centering
    \includegraphics[width=\linewidth]{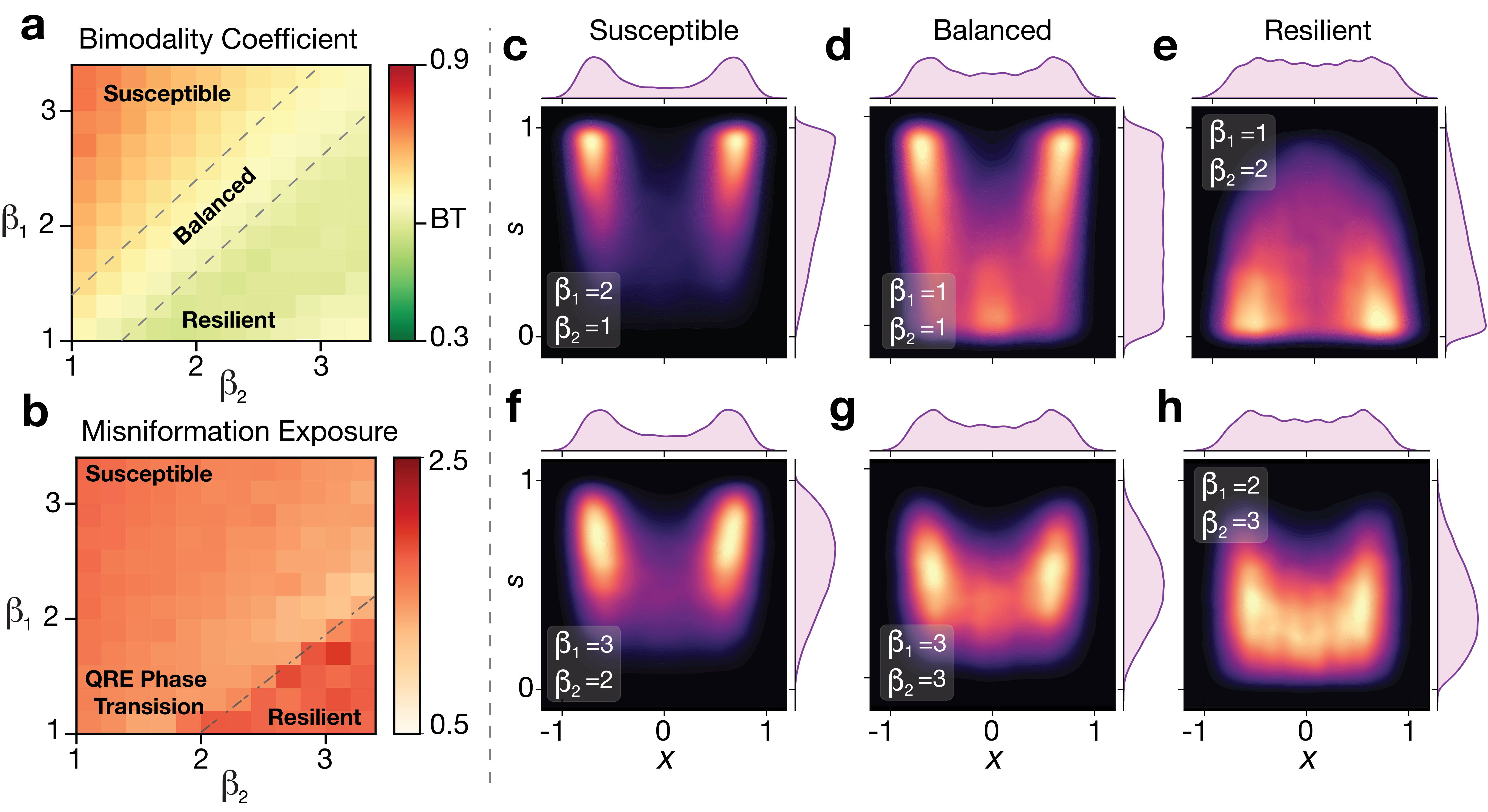}
    \caption{ Decreasing community susceptibility can depolarize a susceptible community. However, it could result in more misinformation exposure for individuals. \textbf{a,b.} bimodality coefficient and misinformation exposure for different susceptibility distributions. \textbf{c-g.} Final opinion distribution induced by equilibrium for different susceptibility distributions, depicted on the right, with $\tau=10$, $\eta=1$, and $\xi=2$. When a community is resilient to misinformation, players find it difficult to polarize the community. However, making a community extremely resilient can backfire, resulting in a phase transition in equilibrium and increased misinformation exposure.} 
    \label{fig:fixed-beta} 
\end{figure*}

\paragraph{Community Susceptibility}

Figure \ref{fig:fixed-beta} shows how the different distributions of individual susceptibility influence the equilibrium and the resulting opinion distribution. As the community becomes more resilient to misinformation, we observe a shift towards depolarization, and misinformation exposure decreases. Reducing the average susceptibility beyond a certain threshold introduces a phase transition in the equilibrium and creates a spike in misinformation exposure. Similar to high credibility gain, players start to exploit low-susceptible users for polarization.

Even in communities with a balanced susceptibility distribution, the equilibrium often remains in phase one. However, in these scenarios, there are enough resilient users who can independently form a central group. This resilient population plays a crucial role in depolarizing the community by preventing players from exploiting susceptible individuals and acting as a communication bridge between the parties.

\section*{Misinformation Interventions}

Approaches to counter misinformation\cite{roozenbeek2023countering}, include debunking\cite{chan2017debunking}, fact-checking\cite{vo2018rise}, source accountability\cite{graves2019information}, and enhancing community resilience to misinformation\cite{van2022misinformation}. Each of these interventions affects specific aspects of the proposed model. Therefore, we can utilize the detailed sensitivity analysis from the previous section to assess the effectiveness of these strategies in countering misinformation and determine when they are most efficient.

\paragraph{Debunking Misinformation}

Fact-checking and debunking, which involve delivering factual information to expose misinformation, are effective tactics for combating misinformation. However, the response of misinformed individuals to such interventions may sometimes be different than expected\cite{diaz2023disinformation}. 
Debunking targets the susceptible part of the community and aims to deliver factual information to fight misinformed individuals. Thus, in our model, increased debunking efforts correlate with a reduction in misinformation gain, represented by $\eta$. According to our results depicted in Figure \ref{fig:xi-eta}, the current media landscape is in phase one, indicating that decreasing misinformation gain is a more efficient tactic than increasing credibility gain.

Debunking requires delivering factual information to community segments heavily exposed to misinformation. Results from Figure \ref{fig:Mixed}(c) show that radical sources primarily influence these segments. However, most debunking efforts originate from opposing and centrist sources\cite{graves2012fact}. Thus, the likelihood of successfully reaching the targeted community is low, making the reduction of misinformation through debunking and fact-checking practically challenging in the current media landscape. Nevertheless, if debunking originates from sources within the same party as the audience, it can effectively reach the misinformed section of the communities, reduce misinformation gain, and shift the equilibrium towards reduced misinformation exposure and polarization.

\paragraph{Source Accountability}

In the proposed model, credibility gain, $\xi$, enforces how individuals resilient to misinformation are affected by information shared from sources with low credibility. Therefore, holding sources accountable for their actions and stressing their credibility and trustworthiness among the public is equivalent to increasing credibility gain\cite{plaisance2000concept,fengler2012media}. Unlike misinformation gain, which primarily affects susceptible parts of the community, credibility gain impacts resilient individuals. Therefore, holding sources accountable can be an effective method to increase credibility gain, especially since the equilibrium in phase one positions low-susceptible individuals centrally, where information about the low credibility of radical sources is plentiful. Given that the current equilibrium corresponds to phase one in Figure \ref{fig:xi-eta}(c), this strategy decreases polarization and reduces average misinformation exposure. Thus, making media outlets and elites accountable is an efficient solution to counter misinformation, particularly for non-partisan entities aiming to shift the equilibrium towards a more informed community.

\paragraph{Community Susceptibility} 

Our findings highlight the critical role of enhancing community resilience in combating misinformation. Figure \ref{fig:fixed-beta} shows how the changing susceptibility of the community can result in depolarized equilibrium with lower levels of misinformation exposure. It is evident that we can substantially alter the information ecosystem by fostering a more resilient community. Additionally, our results suggest that introducing resilient users into the community and encouraging the formation of a low-susceptibility centrist group can be a strong intervention to depolarize the community and compel the players to reduce misinformation exposure.

It is also worth noting that our results emphasize the effectiveness of educational initiatives aimed at improving media literacy and lower susceptibility. By equipping individuals with the skills to better assess the credibility of the information they consume, one can decrease the overall susceptibility to the community. While this is an expensive and long-term solution, it discourages the players from sharing misinformation and diminishes its potential to polarize public opinion. Focusing on empowering individuals with knowledge and critical skills is crucial for fostering a more informed and less divided society.

\paragraph{Arms Race}

The experimental results in Figure \ref{fig:Mixed} reveal a critical vulnerability in the current environment. When one player significantly reduces its credibility below the equilibrium, the optimal response for the opponent shifts to more misinformation dissemination. This dynamic reveals the vulnerability of the current information landscape to excessive misinformation and highlights the ease with which adversaries can manipulate it to escalate misinformation problems. Consider a scenario where an adversary establishes a low-credibility source. Regardless of its political stance, this action disrupts the existing equilibrium. Initially, it undermines the perceived credibility of the affiliated party. Consequently, the opposing party feels compelled to disseminate more misinformation to remain competitive. This results in a lose-lose situation for all players involved in this game.

Conversely, a biased news source with high credibility will improve the equilibrium by prompting competitors to prioritize factual reporting. Promoting or supporting news sources with high credibility, even if they are biased, elevates information quality in the affiliate party and enhances the equilibrium, prompting the other players in the game to compete with high credibility rather than sensationalism.

\section*{Conclusion}

We present a new perspective for studying misinformation by examining the competition over public influence when sharing misinformation is possible. The approach we present sheds light on news sources' decision-making process and potential equilibria behavior they can converge. Our results also provide insights regarding the efficiency of different interventions to combat misinformation.

We draw from the existing social science literature on misinformation and present a detailed and adaptable mathematical formalism to model the interplay between misinformation, news sources, and public opinion, 
The model explains how misinformation dissemination can lead to polarization and investigates the possible parametrizations that permit the players to exploit misinformation to polarize the community.

The proposed approach in this paper opens new avenues for studying the problem of misinformation by focusing on misinformative narratives as long-term dynamic strategies to influence public opinion, instead of singular instances. This shift of focus acknowledges that misinformation is not a one-time occurrence but a persistent issue that evolves with news sources, technologies, and public perception dynamics. By treating misinformation as a dynamic problem, we can apply control and game theory principles to model, predict, and correct the behavior of different actors within this game toward a factual information landscape. This continuous nature of the problem underscores the urgency of a dynamic approach to solving the current misinformation issue.

Several refinements can improve the fidelity of the model we present. For instance, we assume that susceptibility is independent of misinformation exposure. However, in reality, exposure to misinformation can increase susceptibility, further exacerbating misinformation propagation. Another promising direction for future research is allowing users in the model to share misinformation on social networks and investigating how online misinformation interacts with public opinion. These refinements underscore the significance of the proposed model as a tool for informed and safe decision-making in the information ecosystem.


The sensitivity analysis demonstrates that there are two possible phases for the equilibrium. It indicates that a transition is possible if the attention gained through spreading misinformation or the penalty for low-credible media outlets increases. This transition can reduce misinformation exposure and depolarize the community. 
Therefore, the proposed model allows for investigating the effectiveness of various intervention methods for countering misinformation. This analysis provides public policy-makers and governmental entities with the essential information to make informed decisions about combating misinformation.

\bibliography{Ref}
\end{document}